\documentclass[twocolumn,superscriptaddress,amsmath,amssymb,showpacs,showkeys]{revtex4}

\usepackage{graphicx,color}
\usepackage{dcolumn}
\usepackage{bm}
\usepackage[colorlinks=true]{hyperref}
\newcommand{\beq}{\begin{equation}} \newcommand{\eeq}{\end{equation}}
\newcommand{\bqa}{\begin{eqnarray}} \newcommand{\eqa}{\end{eqnarray}}

\definecolor{gold}{rgb}{0.75,0.56,0.00}
\definecolor{green}{rgb}{0.00,0.50,0.00}

\begin{document}

\preprint{APS/123-QED}

\title{Efficient Quantum Filtering for Quantum Feedback Control}

\author{Pierre Rouchon}
 \email{pierre.rouchon@mines-paristech.fr}
  \affiliation{Centre Automatique et Systemes,, Mines ParisTech, PSL Research University,  60 Boulevard Saint-Michel, 75272 Paris Cedex 6, France.}
\author{Jason F. Ralph}
 \email{jfralph@liv.ac.uk}
  \affiliation{ Department of Electrical Engineering and Electronics, The University of Liverpool,\\  Brownlow Hill, Liverpool, L69 3GJ, United Kingdom.}

\date{\today}

\begin{abstract}
We discuss an efficient numerical scheme for the recursive filtering of diffusive  quantum stochastic master equations. We show that the resultant quantum trajectory is robust and may be used for feedback based on inefficient measurements. The proposed numerical scheme is amenable to approximation, which can be used to further reduce the computational burden associated with calculating quantum trajectories and may allow real-time quantum filtering. We provide a two-qubit example where feedback control of entanglement may be within the scope of current experimental systems.
\end{abstract}



\maketitle

\section{\label{sec:sec1}INTRODUCTION}

Experiments have demonstrated the ability to use continuous weak measurements to control the behavior of single quantum systems in a variety of technologies: including optical systems \cite{Arm2002}, atom-optical systems \cite{Say2011}, superconducting microwave systems \cite{Vij2012}, and NV diamond \cite{Blo2014}. More recently, experiments have shown that it is possible to reconstruct quantum trajectories from the experimental records generated by continuous weak measurements in superconducting microwave devices \cite{Mur2013,Web2014}.

The use of continuous weak measurements to probe the behavior of individual quantum mechanical devices has been considered for over 30 years. This is often referred to as {\em quantum filtering}. That is, generating an estimate of the underlying state of the quantum system that is conditioned on the measurements that have been obtained. Belavkin formulated the basic theory for quantum filtering in the early 1980's \cite{Bel1980}, and related approaches evolved separately in the late 1980's in a number of forms, see \cite{Ple1998,Wis2010} for relevant reviews. Using classical signals generated by weak measurements to effect controls is often referred to as measurement-based quantum control \cite{Zha2014} or incoherent quantum feedback control \cite{Don2010}, as opposed to coherent quantum feedback control \cite{Mab2008,Nur2009} -- see \cite{Zha2014} for recent comprehensive review of the subject.

Continuous weak measurements are normally derived by considering a direct (projective) measurement on a coupled quantum system, consisting of an environment and the quantum system of interest. If the environment is much larger than the system of interest, any back action of the projective measurement on the combined system will have a relatively small (i.e. weak) effect on the system of interest, compared to its natural evolution. It is also common to make a Markov approximation, which assumes that any correlations between the system and the environment are damped on timescales shorter than any of the timescales present in the system of interest. By averaging over the environmental degrees of freedom, it is possible to obtain an effective evolution for the system of interest, either in average form (the master equation) or in a form which provides a quantum state that is conditioned on the specific measurement record obtained for the single system (the stochastic master equation, SME). The evolution of the conditional state represents the best estimate of the evolution of the single quantum system of interest and provides a quantum trajectory for that system (often referred to as an `unraveling' of the master equation).

The ability to generate a quantum trajectory from an experimental measurement record provides an opportunity to implement a particular form of measurement-based/incoherent feedback control called {\em rapid purification}. In 2003, Jacobs proposed the use of quantum feedback control to modify the speed at which a completely mixed (unknown) quantum state could be purified \cite{Jac2003}. That is, feedback is used to modify the rate at which information about the quantum state is extracted from the system. In \cite{Jac2003}, Jacobs showed that the use of measurement-based feedback could increase the average rate of purification for a single qubit by up to a factor of two. This control protocol is relevant for future experiments because existing experiments require that the system be initialized in a known pure state using a projective measurement before the quantum trajectory can be found \cite{Mur2013,Web2014}. More general approaches using mixed initial states require information to be extracted gradually as the weak measurements are recorded. Subsequent studies on rapid purification showed that the proposed protocol was optimal, in that it maximized the average rate of purification, but it did not minimize the time taken to reach a given level of purity \cite{Wis2006}. Other work has generalized these results for different optimization conditions \cite{Wis2008}, N-level systems \cite{Com2008}, practical implementation of the controls \cite{Gri2007}, shared entangled states \cite{Hil2007}, imperfections \cite{Com2011}, inefficient detection \cite{Li2013}, and mixed protocols \cite{Li2013,Bal2014}. In particular, Li et al. demonstrated that when the efficiency of the detector is lower than 50\%, there is no predicted speed up in purification rate \cite{Li2013}. This result is significant for the recent quantum trajectory experiments in superconducting microwave systems because the estimated detection efficiency is given as 40\% in \cite{Web2014}.

The paper has two main aims: the extension of a numerical integration method \cite{Ami2011} to address the case of inefficient continuous weak measurements, and the application of this method to a specific example of quantum control for two coupled qubits in a mixed initial state. The paper provides an efficient numerical method to solve the stochastic master equation for a given measurement record or realization. It uses this method to demonstrate a two-qubit feedback control scheme for completely mixed initial states that is robust and amenable to implementation in current experimental systems. The integration method was discussed for efficient continuous weak measurements in \cite{Ami2011}; here it is adapted to deal with inefficient detection and extended to include higher order corrections. The resultant method ensures the positiveness of the conditional quantum state (as represented by the density matrix) and can be simplified further, reducing the computational load associated with the solution of the SME without a commensurate effect on the accuracy of the conditioned state. The intention is to develop a quantum filter that could be used for online quantum state estimation and feedback control.

\section{\label{sec:sec2}NUMERICAL METHOD}

For a quantum system, represented by a density matrix $\rho$, coupled to a Markovian environment, represented by a set of environmental operators $V_j$  $(j = 1\dots m)$, the master equation in Linblad form is given by \cite{Wis2010},
\begin{equation}\label{me1}
d\rho = -i\left[H,\rho\right]dt +\sum_{j=1}^{m} \left\{ V_{j} \rho V^{\dagger}_{j} -\frac{1}{2}\left(V^{\dagger}_{j} V_{j} \rho + \rho V^{\dagger}_{j} V_{j} \right)\right\}dt
\end{equation}
where $H$ is the Hamiltonian of the system and $dt$ is an infinitesimal time increment.

If we select some of the environmental operators to provide a set of continuous weak measurements $L_r$ $(r = 1\dots m', m' < m)$, with efficiencies $\eta_r$, the corresponding stochastic master equation for the conditioned state $\rho_c$ is then given by \cite{Wis2010},
\begin{eqnarray}\label{sme1}
d\rho_c&=&- i \left[H,\rho_c\right]dt \nonumber\\
&&+\sum_{j=1}^{m-m'} \left\{ V_{j} \rho_c V^{\dagger}_{j} -\frac{1}{2}\left(V^{\dagger}_{j} V_{j} \rho_c + \rho_c V^{\dagger}_{j} V_{j} \right)\right\}dt  \nonumber\\
&&+\sum_{r=1}^{m'} \left\{ L_{r} \rho_c L^{\dagger}_{r} -\frac{1}{2}\left(L^{\dagger}_{r} L_{r} \rho_c + \rho_c L^{\dagger}_{r} L_{r} \right)\right\}dt   \nonumber\\
&&+ \sum_{r=1}^{m'} \sqrt{\eta_r}\left(L_{r}\rho_c+\rho_c L^{\dagger}_{r}-\mathrm{Tr}(L_{r}\rho_c+\rho_c L^{\dagger}_{r}) \right)dW_{r}\nonumber\\
\end{eqnarray}
where we have taken $dW_{r} $ to be a real Wiener increment (such that $dW_{r}=0$ and $dW_{r} dW_{r'}  = \delta_{rr'}dt$). The measurement record for each of the measurement operators $L_{r}$ during a time step $t\rightarrow t+dt$ is given by, $dy_r(t)= \sqrt{\eta_j}\mathrm{Tr}(L_{r}\rho_c+\rho_c L^{\dagger}_{r}) dt+dW_{r}$. The previous case considered in \cite{Ami2011} deals with the case where $\eta_r = 1$ for all environmental operators $L_r$.

For a given initial state, a system and a set of environmental operators, the SME may be integrated using standard numerical stochastic integration methods, e.g. the Euler-Maruyama method (which is weakly convergent to first order) or the Euler-Milstein method (strongly convergent to first order) \cite{Mil1995}. For an SME of the form (\ref{sme1}) with commuting measurement operators  ($[L_{r},L_{s}] =0$ for $r,s$ between $1$ and $m'$),  the Euler-Milstein increment for a finite time step $\Delta t$ and finite stochastic increments $\Delta W_j$ can be computed using~\cite{Mil1995,Tal1995},
\begin{widetext}
\begin{eqnarray}\label{sme2}
\Delta \rho_c&=&- i \left[H,\rho_c\right]\Delta t+\sum_{j=1}^{m-m'} \left\{ V_{j} \rho_c V^{\dagger}_{j}
-\frac{1}{2}\left(V^{\dagger}_{j} V_{j} \rho_c + \rho_c V^{\dagger}_{j} V_{j} \right)\right\}\Delta t   \nonumber\\
&&+\sum_{r=1}^{m'} \left\{ L_{r} \rho_c L^{\dagger}_{r} -\frac{1}{2}\left(L^{\dagger}_{r} L_{r} \rho_c + \rho_c L^{\dagger}_{r} L_{r} \right)\right\}\Delta t
+ \sum_{r=1}^{m'}  \sqrt{\eta_r}\left(L_{r}\rho_c+\rho_c L^{\dagger}_{r}-\mathrm{Tr}(L_{r}\rho_c+\rho_c L^{\dagger}_{r}) \right)\Delta W_{r}  \nonumber \\
&&+ \sum_{r,s=1}^{m'}  \frac{\sqrt{\eta_r \eta_s}}{2}
\left(
\begin{array}{l}
L_r L_s \rho_c+\rho_c L^{\dagger}_{r} L^{\dagger}_{s}+L_{s}\rho_c L^{\dagger}_{r}+ L_{r}\rho_c L^{\dagger}_{s} \\
-\mathrm{Tr}\left(
\begin{array}{l}
L_r L_s \rho_c
+\rho_c L^{\dagger}_{r} L^{\dagger}_{s}+L_{s}\rho_c L^{\dagger}_{r}+ L_{r}\rho_c L^{\dagger}_{s}
\end{array}\right)\rho_c \\
-\mathrm{Tr}(L_s\rho_c+\rho_c L^{\dagger}_{s}) (L_r\rho_c+\rho_c L^{\dagger}_{r})\\
-\mathrm{Tr}(L_r\rho_c+\rho_c L^{\dagger}_{r})(L_s\rho_c+\rho_c L^{\dagger}_{s})\\
+2\mathrm{Tr}(L_r\rho_c+\rho_c L^{\dagger}_{r})\mathrm{Tr}(L_s\rho_c+\rho_c L^{\dagger}_{s}) \rho_c
\end{array}
\right)\left(\Delta W_r\Delta W_s-\delta_{r,s}\Delta t\right)
\end{eqnarray}
\end{widetext}
where $\delta_{r,s}$ is the Kronecker delta symbol, and the $\Delta W_r$'s are independent Gaussian variables with zero mean and a variance equal to $\Delta t$.

The Euler-Milstein increment is strongly convergent to first order in $\Delta t$, but the stochastic increment $\Delta W_r$ is only half order in $\Delta t$, so $(\Delta W_r)^2$ terms need to be retained \cite{Mil1995}. The form given in (\ref{sme2}) does guarantee that $\rho_c$ is Hermitian, up to numerical rounding errors due to the accuracy of the floating point calculations, however, it is computationally expensive and the formula does not guarantee the positivity of the resultant density matrix. An alternative form for the increment, which ensures the positivity of $\rho_c$ can be derived from an alternative expansion~\cite{Ami2011} to second order in $\Delta W_r$ that provides a positive operator valued measure (POVM) \cite{Wis2010}. The increment for a time step $t=n\Delta t\rightarrow (n+1)\Delta t$ is then given by, $\Delta \rho_c(n) = \rho_c(n+1)- \rho_c(n)$, where
\begin{widetext}
\begin{equation}\label{sme3}
\rho_c(n+1)= \frac{M_n\rho_c(n)M_n^{\dagger}+\sum_{j=1}^{m-m'}V_j\rho_c(n)V_j^{\dagger}\Delta t  + \sum_{r=1}^{m'}(1-\eta_r)L_r\rho_c(n)L_r^{\dagger}\Delta t }
{\mathrm{Tr}\left(M_n\rho_c(n)M_n^{\dagger}+ \sum_{j=1}^{m-m'}V_j\rho_c(n)V_j^{\dagger}\Delta t + \sum_{r=1}^{m'}(1-\eta_r)L_r\rho_c(n)L_r^{\dagger}\Delta t \right)}
\end{equation}
and where the operator $M_n$ is given by
\begin{eqnarray}\label{Mn1}
M_n &=& I-\left(iH + \frac{1}{2}\sum_{j=1}^{m-m'} V^{\dagger}_{j}  V_{j} +\frac{1}{2}\sum_{r=1}^{m'} L^{\dagger}_{r}L_{r}\right)\Delta t  \nonumber\\
&&+\sum_{r=1}^{m'} \sqrt{\eta_r}L_{r}\left(\sqrt{\eta_r}\mathrm{Tr}(L_{r}\rho_c(n)+\rho_c(n)L^{\dagger}_{r})\Delta t +\Delta W_{r} (n)\right)
+\sum_{r,s=1}^{m'}  \frac{\sqrt{\eta_r \eta_s}}{2}L_{r}L_{s}(\Delta W_r(n)\Delta W_s(n)-\delta_{r,s}\Delta t)
\end{eqnarray}
\end{widetext}
However, we note that the integrated measurement records over the time step $\Delta y_r(n)$ are given by,
\begin{equation}\label{meas}
\Delta y_r(n) = \sqrt{\eta_j}\mathrm{Tr}(L_{r}\rho_c (n)+\rho_c(n) L^{\dagger}_{r}) \Delta t+\Delta W_{r}(n)
\end{equation}
so the operator $M_n$ can be simplified to give
\begin{widetext}
\begin{equation}\label{Mn2}
M_n = I-\left(iH + \frac{1}{2}\sum_{j=1}^{m-m'} V^{\dagger}_{j}  V_{j} +\frac{1}{2}\sum_{r=1}^{m'} L^{\dagger}_{r}L_{r}\right)\Delta t
+\sum_{r=1}^{m'} \sqrt{\eta_r}L_{r}\Delta y_r(n)+\sum_{r,s=1}^{m'}  \frac{\sqrt{\eta_r \eta_s}}{2}L_{r}L_{s}(\Delta y_r(n)\Delta y_s(n)-\delta_{r,s}\Delta t)
\end{equation}
\end{widetext}
where the state update is still given by (\ref{sme3}).

\section{\label{sec:sec3}EXAMPLE SYSTEM - TWO COUPLED QUBITS}

To demonstrate the efficiency and the robustness of the proposed method, we consider a specific two-qubit example. A two qubit example has been selected because single qubit quantum control and state estimation has been studied extensively \cite{Jac2003,Wis2006,Wis2008,Gri2007,Com2011,Li2013,Bal2014}, and two-qubit quantum control offers the potential to manipulate the entanglement between the two qubits as well as the rate of purification \cite{Hil2007,Car2007,Car2008,Mas2010}. Examples demonstrating that quantum feedback can be used to stabilise or control the level of entanglement in two qubit systems have been discussed for stochastic quantum jump processes \cite{Car2007,Car2008,Mas2010} and continuous unravelings of the master equation \cite{Hil2007,Car2008,Mas2010}.

We consider the case where each qubit has an intrinsic Hamiltonian $H_0 = \omega X/2$, where $X$, $Y$ and $Z$ are the Pauli matrices. This Hamiltonian generates a rotation of the individual qubit Bloch vector around the respective $X$-axis, with angular frequency $\omega$. We introduce a coupling between the two qubits that couples the $Z$-components of the Bloch vectors. The total Hamiltonian (in the absence of controls) is given by,
\begin{equation}\label{H}
H=\frac{\omega}{2}(XI + IX)+\kappa (ZZ)
\end{equation}
where the tensor product is implied (i.e. $ZZ=Z\otimes Z$), $I$ is the identity matrix, and $\kappa$ is the strength of the coupling between the two qubits.

In most of the cases considered here, we will fix the qubit rotation frequency to be a standard value, $\omega = \omega_0$, which will be used to fix the size of the time steps used in the numerical integration, so that $\Delta t = 2\pi/(N\omega_0)$ and $N$ is the number of steps per cycle. Each of the qubits is measured along its Bloch $Z$-axis and these measurements are inefficient, so that the environmental operators are $L_{1}= \sqrt{2 k_1} (ZI)$ and $L_{2}=\sqrt{2 k_2} (IZ)$, where $k_1$ and $k_2$ are the measurement strengths, with efficiencies $\eta_1$ and $\eta_2$ respectively. We will also fix the qubit-qubit coupling and the measurement strengths to be relatively weak so that the individual qubit evolution is perturbed but not dominated by the coupling or the measurement interactions.  We use $\kappa = 0.010\omega_0$, $k_1=k_2=0.005\omega_0$, and $\eta_1=\eta_2 = 0.85$ in all of the examples. The measurement efficiencies are better than those available to the experimental systems described in \cite{Mur2013,Web2014}; however, the values are not so close to 100\% efficiency so as to be unrealistic. Lower values for the measurement efficiency do reduce the level and robustness of the steady state entanglement found in the two-qubit control example below, but the results presented below are not critically dependent on the precise value of the measurement efficiency. The two (local) measurement records are
\begin{eqnarray}\label{Dy}
\Delta y_1 &=&  \sqrt{8 k_1 \eta_1}Tr(ZI \rho)+\Delta W_1  \nonumber \\
\Delta y_2 &=&  \sqrt{8 k_2 \eta_2}Tr(IZ \rho)+\Delta W_2   \nonumber
\end{eqnarray}
and there are no extraneous unprobed environmental degrees of freedom, $V_{j}$, for simplicity.

\section{\label{sec:sec4}NUMERICAL ACCURACY}

To explore the accuracy of the proposed method, a separable initial pure state and a completely mixed (unknown) initial state are integrated over 50 oscillation periods using the Euler-Milstein increment with $\omega = \omega_0$ and $N=5000$ steps per cycle. No controls are applied, and the qubits are allowed to rotate freely around their respective $X$ axes. Figure 1a shows the average purity for both the pure state and the mixed state -- both states converge to an entangled mixed state with an average purity of around 0.85. The integrated state found using 5000 steps per cycle is then compared against integrated states found using the Euler-Milstein method with longer integration steps (i.e. fewer steps per cycle) and compared against the proposed method with much longer time steps. The only modification to the methods described above is due to the fact that the accuracies of the resultant states are quite sensitive to the Hamiltonian part of the integration step. Because of this sensitivity, the Hamiltonian evolution terms are expanded to at least second order in $\Delta t$. The other second order terms in $\Delta t$ are much smaller and can be removed). The comparison uses the Fidelity, $F$ \cite{Pet2004},
$$
F= F(\rho_{0},\rho_{c}) = \left|{\rm Tr}\left[\sqrt{\sqrt{\rho_{c}} \rho_{0} \sqrt{\rho_{c}}}\right]\right|^2
$$
between the estimated (conditioned) state and the conditioned state generated using 5000 steps/cycle, $\rho_0$, as a metric.
\begin{figure}[htbp]\label{Fig_1}
	\centering
		\includegraphics[width=1.05\hsize]{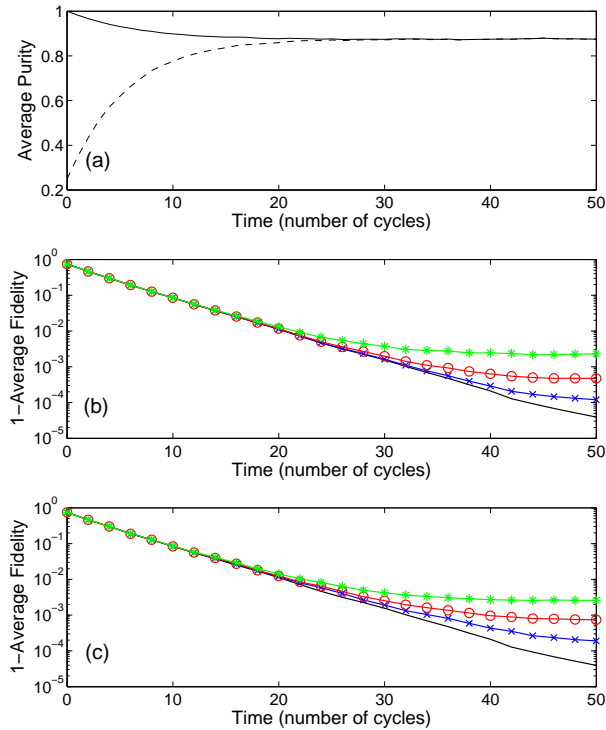}
	\caption{\label{fig:Average Fidelity} (Color online) (a) Average Purity for initial pure state (solid black line) and initial completely mixed state (dashed black line) calculated using the Euler-Milstein method with 5000 steps per cycle; (b) 1-Average Fidelity, where Fidelity is calculated between an initial pure state with 5000 steps per cycle and an initial completely mixed state using Euler-Milstein method with 1000 steps (blue-cross), 500 steps (red-circle) and 250 steps (green-asterisk) per cycle; and, (c) 1-Average Fidelity, where Fidelity is calculated between initial pure state with 5000 steps per cycle and an initial completely mixed state using the proposed method with 250 steps (blue-crosses), 100 steps (red-circles) and 50 steps (green-asterisks) per cycle -- all values averaged over 1000 realizations.}
\end{figure}

Figure 1b shows the accuracy of the Euler-Milstein integration method for different number of steps per qubit cycle: 1000 steps/cycle, 500 steps/cycle and 250 steps/cycle, and gives the average value of $1-F$, where the fidelity is averaged over a large number of realizations (1000 realizations, in this case). The integration errors reduce with more steps/cycle, as expected, but 250 steps/cycle are sufficient to produce a relatively high fidelity state $F > 99\%$. The corresponding calculations for the proposed method are shown in Figure 1c, and also compared against the state generated using the Euler-Milstein method with 5000 steps/cycle. In Figure 1c, the number of steps per cycle are 250, 100 and 50. The same accuracy can be achieved using the state update (4) with 4-5 times fewer integration steps than the standard Euler-Milstein method: 50 steps per cycle provides a conditioned state with high average fidelity $F > 99\%$. Recent experiments have generated quantum trajectories using measurements that had been sampled around 60 times per qubit oscillation period \cite{Web2014}, so measurement rates of around 50-60 measurements per cycle can be considered to be the current `state-of-the-art' for practical systems, albeit without feedback control. To move towards a practical implementation of quantum feedback control in such systems, the quantum trajectories and the corresponding controls would need to be generated in real time.

Although it is not shown here, the numerical benefit is even more pronounced when the purity of the state approaches one. The Euler-Milstein method can give rise to unphysical conditioned states with purities significantly above one. This numerical instability is more with larger integration steps, and it can lead to unreliable results numerical integration. For the proposed method, the normalization contained in the POVM-like update (4) dramatically reduces the likelihood of such unphysical conditioned states and the resultant purity is significantly more stable than that found using the Euler-Milstein method.

\section{\label{sec:sec5}EXAMPLE - QUANTUM CONTROL}

In this section, we examine how the proposed numerical method can be used with quantum feedback control to affect the entanglement of the two qubits described by the Hamiltonian (\ref{H}). The two measurement records are local and we restrict ourselves to local controls, for simplicity and to demonstrate that feedback control provides some advantages even without explicit control over the entangling interaction itself \cite{Car2007,Car2008,Mas2010}. In the one-qubit case, the approach to quantum rapid purification has centered on the use of Bloch rotations to rotate the conditioned qubit state towards the plane orthogonal to the measurement axis \cite{Jac2003}, or towards the measurement axis \cite{Wis2006}, or some combination of the two \cite{Li2013,Bal2014}. These approaches have the advantage of being relatively robust and insensitive to small errors in these rotations \cite{Gri2007}. In the two qubit case, we examine the use of local controls that rotate the Bloch vectors for the individual qubits (corresponding to a partial trace of the density matrix over the other qubit state) towards one or other of the three main axes: i.e. rotations towards the $XI$, $YI$ or $ZI$ axis for qubit 1, and $IX$, $IY$ or $IZ$ axis for qubit 2. The controls applied are calculated at each time step by taking the estimated one qubit Bloch vector and a target state (aligned along one or other of the one qubit axes) and then rotating the Bloch vector onto the target state using a rotation based on Rodriques' rotation formula \cite{Che1989}. The resultant rotations are similar to bang-bang controls \cite{Vio1998}, but once the Bloch vectors are close to the target axes, the controls are relatively small. The controls are applied at the end of an integration time step, so the number of time steps reflects the number of control rotations that are required in every qubit cycle.

We characterize the entanglement in terms of the concurrence \cite{Woo1998} and the negativity \cite{Wei2003}. Two entanglement metrics are used because the ordering of mixed states in terms of their entanglement (i.e. the entanglement of the state $\rho$ is greater than that of the state $\rho'$) can be different for different entanglement measures \cite{Wei2003}, and two different metrics gives more confidence that the relative ordering of the resultant states is robust -- although, in some of the examples, the difference between the concurrence and the negativity is very small and only the concurrence values will be given. For a two-qubit mixed state density matrix $\rho$ the concurrence is given by
$$
C(\rho)=max\{0,\sqrt{\lambda_1}-\sqrt{\lambda_2}-\sqrt{\lambda_3}-\sqrt{\lambda_4}\}
$$
where $\sqrt{\lambda_1},\ldots,\sqrt{\lambda_4}$ are the eigenvalues of the matrix $\rho(YY)\rho^*(YY)$ in nondecreasing order and the negativity $\mathcal{N}$ is given by twice the absolute sum of the negative eigenvalues of the partial trace of $\rho$ with respect to one of the individual systems \cite{Wei2003}.

\begin{figure}[htbp]\label{Fig_2}
	\centering
		\includegraphics[width=1.05\hsize]{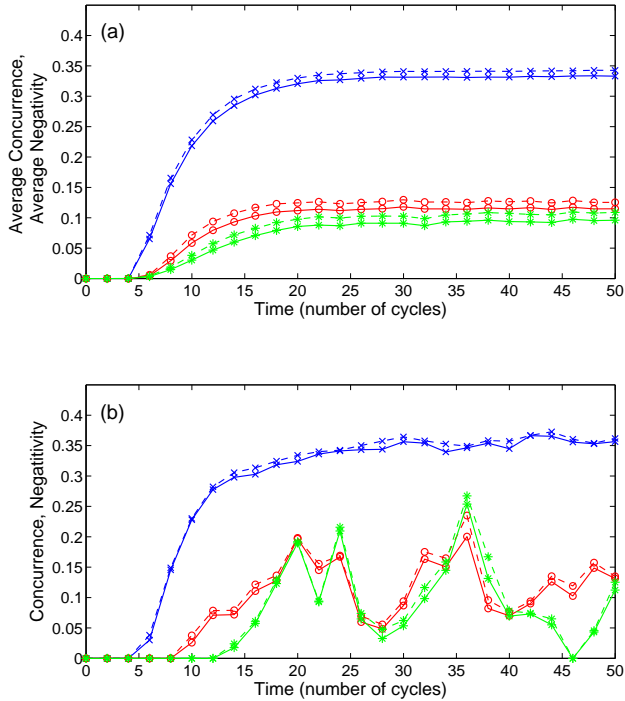}
	\caption{\label{fig:Entanglement} (Color online) (a) Average Concurrence (dashed lines) and average Negativity (solid lines) for the three control strategies for a completely mixed initial state, averaged over 500 realizations and 250 steps per cycle using the proposed integration scheme: two-qubit $YI$ \& $IY$ controls (blue-crosses), one-qubit $XI$ controls (red-circles), and no controls/free rotation about the qubit $X$ axes (green-asterisks); (b) Concurrence (dashed lines) and Negativity (solid lines) for one realization of the three control strategies in (a).}
\end{figure}

If no controls are applied to the system, and the qubits are allowed to rotate freely under the action of their respective Hamiltonians, there is small build up of entanglement in the system because of the $ZZ$ coupling term. Figure 2a shows the average entanglement for a completely mixed initial state with no controls applied (green lines with asterisks). The average concurrence is limited to around $C\simeq 0.11-0.12$ ($\mathcal{N} \simeq 0.09-0.10$). Figure 2b shows a single realization, where the entanglement varies significantly as a function of time (between zero and $C\simeq 0.20$). Figure 2 also shows the average entanglement if controls are only applied one of the two qubits: one qubit Bloch vector is rotated towards its $X$ axis ($XI$, in this case) and the other is allowed to rotate under the action of the Hamiltonian. The result is a small improvement in the rate of generation of entanglement and a marginal improvement in the average entanglement value $C\simeq 0.12-0.13$ ($\mathcal{N} \simeq 0.11-0.12$), and -- as with the no control case -- the entanglement for an individual realization is stochastic. Controls that rotate one of the qubits towards one of the other axes, $Y$ or $Z$, do not generate entanglement as long as the intrinsic Hamiltonian rotations are significant.

Due to the symmetry of the example (two identical qubits), there are six possible two-qubit controls available, rotating towards the following pairs of axes $XI/IX$, $XI/IY$, $XI/IZ$, $YI/IY$, $YI/IZ$ and $ZI/IZ$. Of these, only simultaneously rotating towards the $YI$ and $IY$ axes generates a reasonable level of entanglement. The steady state entanglement for the $YI/IY$ controls is approximately $C\simeq 0.34$ ($\mathcal{N} \simeq 0.33$) with small fluctuations around this value, see Figure 2b. (The maximum possible entanglement achievable in this system -- with $\kappa = 0.010\omega_0$ and perfect 100\% measurement efficiencies -- corresponds to $C\simeq 0.45$). Compared with the no control and one qubit control cases, this entanglement is relatively robust -- the stochastic fluctuations are small compared to the average entanglement. The entanglement for the two-qubit $YI/IY$ control case corresponds to a mixed state which is approximately described by
$$
\rho_{YI/IY}\simeq\frac{1}{4}(0.85(YI+YY+IY)-0.40(XZ+ZX))
$$
with smaller $XY$ and $YX$ cross terms. The other combinations provide zero entanglement. It is the combined effect of the rotations about the $X$-axis and the repeated rotations to the individual $YI$ and $IY$ axes that allows entanglement to be generated through the addition of the $XZ$ and $ZX$ terms in $\rho_{YI/IY}$. Note that rotating both qubits towards the $X$ axes (i.e. $XI/IX$ controls) generates zero entanglement, even though rotating one of the two qubits towards its $X$ axis does provide a small amount of additional entanglement.
\begin{figure}[htbp]\label{Fig_3}
	\centering
		\includegraphics[width=1.05\hsize]{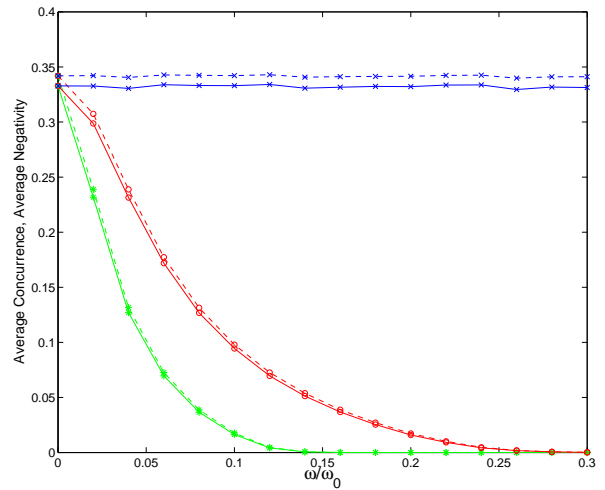}
	\caption{\label{fig:Entanglement2} (Color online) Average Concurrence (dashed lines) and average Negativity (solid lines) for the two-qubit controls rotating towards the $YI/IY$ (blue-crosses), $XI/IY$ (red-circles) and $XI/IX$ (green-asterisks) axes as functions of the Hamiltonian rotation (angular) frequency $\omega/\omega_0$, after 50 cycles ($\omega_0$) averaged over 500 realizations and 250 steps per cycle using the proposed integration scheme.}
\end{figure}

If we were to remove the Hamiltonian evolution (rotation around the respective $X$ axes) the cases $XI/IX$, $YI/IY$ and $XI/IY$ should be identical, and this is indeed the case. However, in these cases -- with no Hamiltonian evolution -- the local controls generate the same level of entanglement as the rotating $YI/IY$ case ($C\simeq 0.34$). Any combination that includes rotations towards one of the individual $Z$ axes generates no entanglement. If rotating the individual Bloch vectors towards the $XI/IY$ or $XI/IX$ axes generates a reasonable level of entanglement in the absence of Hamiltonian rotations, and zero entanglement when $\omega = \omega_0$, this could provide a mechanism to control the level of steady state entanglement in the system. Varying the rate of Hamiltonian rotation allows the build up of entanglement to be moderated without changing the control protocol -- see Figure 3.

\section{\label{sec:sec6}EXAMPLE - APPROXIMATE QUANTUM CONTROL}

The SME provides a best (Bayesian) estimate of the quantum trajectory of the individual system {\em given} the particular continuous weak measurement record or realization, $\Delta y(n)$. The potential problem is that the calculations required to construct this estimate could be impractical to perform in real time, or -- preferably -- quicker than real time to allow feedback controls to be applied in real time. There are a number of ways in which a state estimation filter can be modified to improve the computational efficiency. In classical state estimation, it is common to introduce approximations into the state update (measurement) and state prediction (system dynamics) processes \cite{Bla1986,Bar2001,Ral2010}. The rationale for these approximations is that a state estimate does not necessarily have to be optimal to be useful. In the case of quantum feedback control, the requirement is that the estimated state is sufficiently close to optimal for the quantum filter/SME to be stable under the action of the resultant controls (that is, deviations between the optimal and the approximate state estimate vanish in the long time limit). Standard approaches that could be used to simplify the SME include: (i) removal of small terms, (ii) use of fixed or filtered values to replace the rapidly fluctuating dynamical variables, (iii) analytic approximations, and (iv) discrete values to replace continuous variables (analog-to-digital conversion).

Using the state update (\ref{sme3}), there are a number of benefits in terms of computational efficiency compared to the standard Euler-Milstein method. The first is that the measurement record can be used directly in (\ref{Mn2}) and the majority of the operations applied to the density matrix are constant from time step to time step (assuming that the measurement/environment operators do not change in time). This reduces the number of calculations required to find the new conditioned state by reducing the number of operations that require quantities to be calculated from $\rho_c$. The only operation requiring properties of $\rho_c$ to be calculated is in the normalization step, i.e. division by the trace of the numerator in (\ref{sme3}). In fact, because the positiveness of the density matrix is ensured by the modified update step, the resultant quantum filter is also more stable numerically than the standard Euler-Milstein method, leading to the use of larger time steps seen in Section~\ref{sec:sec4}.

In addition to these numerical benefits, we now consider two approximations to reduce the computational demands even further for the specific control case considered here. The next step is to identify the smallest terms in the $M_n$ operator to truncate the calculations (i.e. removal of small terms). As already stated, the Hamiltonian evolution is dominant, for $\omega \simeq \omega_0$, and the expansion of the Hamiltonian terms needs to be at least second order in $\Delta t$. Of the other terms, the $L^{\dagger}_r L_r\Delta t$ and $(\Delta y_r \Delta y_s-\delta_{r,s})\Delta t$ terms are two orders of magnitude smaller than the other terms, and they can be removed without a significant effect on the fidelity of the final estimated state. The resultant approximate state update is then given by,
\begin{widetext}
\begin{equation}\label{sme4}
\rho_c(n+1)= \frac{M_n^{(approx)} \rho_c(n)(M_n^{(approx)})^{\dagger}+(1-\eta_1)2k_1 (ZI)\rho_c(n)(ZI) \Delta t +(1-\eta_2)2k_2 (IZ)\rho_c(n)(IZ)\Delta t }
{\mathrm{Tr}\left(M_n^{(approx)} \rho_c(n)(M_n^{(approx)})^{\dagger}+(1-\eta_1)2k_1 (ZI)\rho_c(n)(ZI) \Delta t +(1-\eta_2)2k_2 (IZ)\rho_c(n)(IZ)\Delta t\right)}
\end{equation}
where
\begin{equation}\label{Mn3}
M_n^{(approx)} = I-\left(iH+(k_1+k_2) I\right)\Delta t  - \frac{1}{2}H^{2}(\Delta t)^{2}
+\sqrt{2 \eta_1 k_1} (ZI)\Delta y_1(n)+\sqrt{2 \eta_2 k_2} (IZ)\Delta y_2(n)\end{equation}
\end{widetext}
The effect of these approximations is to limit the accuracy of the integration and the resultant conditioned state. For 50 steps per cycle, the fidelity of the conditioned state is around 99.5\%. For 20 steps per cycle, the fidelity of the conditioned state is around 98\%; however, the entanglement of the estimated state is robust to the use of the approximate operator (\ref{Mn3}) as long as the number of steps is greater than 20 per cycle. At 10 steps per cycle, there is a significant reduction in the entanglement of the estimated state.
\begin{figure}[htbp]\label{Fig_4}
	\centering
		\includegraphics[width=1.05\hsize]{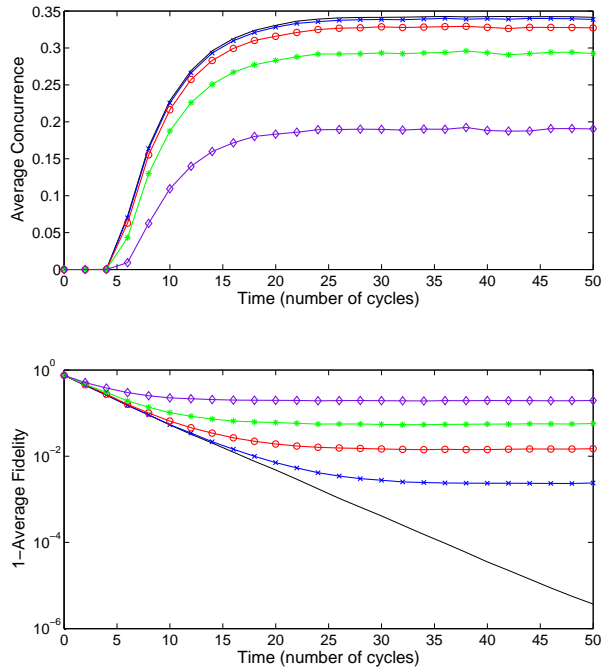}
	\caption{\label{fig:Entanglement3} (Color online) (a) Average Concurrence for the two-qubit controls rotating towards the $YI/IY$ for $\omega = \omega_0$ using the approximate update operator (\ref{Mn3}) with 500 steps/cycle (black-solid line), 50 steps/cycle (blue-crosses), 20 steps/cycle (red-circles), 10 steps/cycle (green-asterisks) and 5 steps/cycle (purple-diamonds); (b) 1-Average Fidelity values corresponding to the concurrence values in (a) -- all values averaged over 500 realizations with fidelities compared against solutions using the proposed integration scheme with 500 steps per cycle.}
\end{figure}
\begin{figure}[htbp]\label{Fig_5}
	\centering
		\includegraphics[width=1.05\hsize]{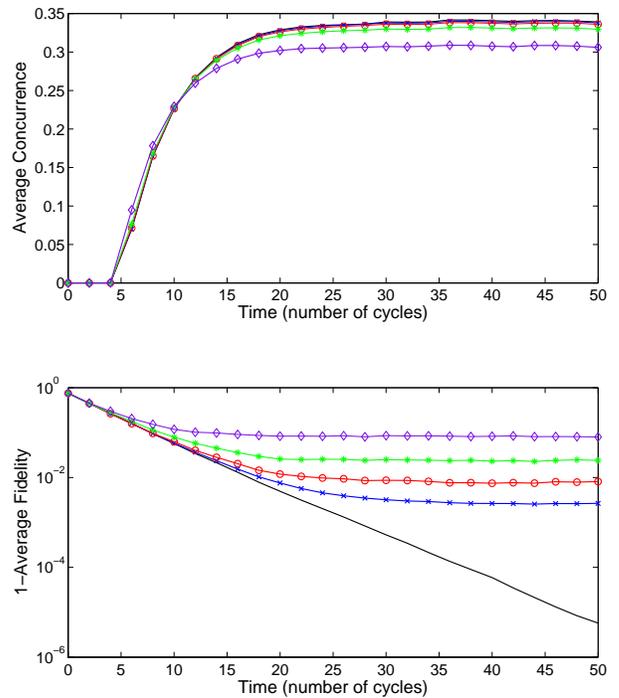}
	\caption{\label{fig:Entanglement4} (Color online) (a) Average Concurrence for the two-qubit controls rotating towards the $YI/IY$ for $\omega = \omega_0$ using a full measurement record with 500 steps/cycle (black-solid line), and 50 steps/cycle with a measurement record truncated to 6 bits per measurement (blue-crosses), 4 bits per measurement (red-circles), 3 bits per measurement (green-asterisks), and 2 bits per measurement  (purple-diamonds); (b) 1-Average Fidelity values corresponding to the concurrence values in (a) -- all values averaged over 500 realizations with fidelities compared against solutions using the proposed integration scheme with 500 steps per cycle.}
\end{figure}

The next approximation that we consider is the precision of the measurement record itself. Any numerical method using an experimental measurement requires that the analog measurements be digitized to a finite number of bits. The ability to digitize an analog measurement is limited by the speed of the analog-to-digital converters used in the experimental set-up, and any saving in terms of the number of bits used to store the measured values will have a knock on effect on the efficiency of the processing. Conventionally, the number of bits used for an experimental value is either 8, 16 or possibly 32. Here, we consider far fewer bits per measurement (sample) value. Figure 5 shows the average Concurrence and $1-$Average Fidelity for different numbers of bits using the approximate operator (\ref{Mn3}): 6 bits, 4 bits, 3 bits and 2 bits. For each number of bits, the signal is discretized between a set of thresholds ($n$ bits, $2^n$ discrete values requires $2^n-1$ threshold values) where the thresholds are uniformly distributed between $\pm 3\sqrt{\Delta t}$. The measurement signals contain a significant Gaussian noise term, so this range corresponds roughly to 3 standard deviations. For 4 or more bits per measurement, the accuracy of the resultant state is unaffected by the reduced number of bits, with the average Fidelity approaching 99\%, and the entanglement of the estimated state is robust for 3 or more bits per measurement. Similar calculations using 20 integration steps per cycle show similarly robust results, with an average Fidelity of around $98\%$. Numerical calculations using a single bit per measurement have been demonstrated elsewhere \cite{Ral2011}, but this requires significantly higher measurement frequencies than the results demonstrated here.

\section{\label{sec:sec7}CONCLUSIONS}

In this paper, we have discussed a method for the efficient integration of the Stochastic Master Equation for a quantum system under the action of continuous weak measurements, and we have demonstrated that it offers significant advantages in terms of accuracy and stability over the Euler-Milstein integration method for finite time steps. We have used an example with two coupled qubits subject to inefficient local measurements and local controls to show that the proposed method can be used with incoherent (measurement-based) quantum feedback control. The fidelity of the estimated (conditioned) quantum state and the resultant quantum trajectory are very good, even when the number of integration steps is very low (or the time steps are very long). In addition, the entanglement between the two qubits is remarkably robust for very low numbers of time steps, and with a number of additional simplifying approximations for the state update process and for the digitization of the measurement signal. Using as few as 3 or 4 bits per measurement and 50 measurements per qubit period would allow the construction of a conditioned state that is sufficiently accurate (fidelity, $F\simeq 99\%$) to allow feedback control to be used to generate states with a concurrence very close to the maximum obtainable.

The ability to use local measurements, an efficient state estimation process, and local quantum feedback controls to manipulate the entanglement between two coupled qubits offers a range of experimental realizations. The required continuous measurement frequencies are within the scope of recent experimental developments and have been demonstrated through the calculation of quantum trajectories from continuous weak measurements \cite{Mur2013,Web2014}. The main obstacles to demonstrating such feedback control of entanglement are the efficiency of the local measurement interactions and the the ability to calculate the updates in real time. The results given above are not critically dependent on the efficiency of the measurement process, but lower efficiency measurements limit the purity and the entanglement found in the final conditioned state. The ability to use fairly coarse approximations in the update step and limited bit digital measurements should assist greatly when implementing the state estimation process and the feedback controls for hardware and/or software reconfigurable processors.

\begin{center}
	\textit{Acknowledgments}
\end{center}
\noindent  The authors would like to thank the Newton Institute (University of Cambridge) for their hospitality, and the organizers of the Workshop on Quantum Control Engineering (July-August 2014). PR would like to thank Claude Le Bris for useful discussions and advice concerning the numerical scheme.

\bibliographystyle{apsrev}

\end{document}